%% file: tba.tex
\documentclass[prl,twocolumn,showpacs,superscriptaddress,preprintnumbers,floatfix]{revtex4}

\usepackage{ifpdf}
\usepackage{hyperref}
\usepackage{dcolumn}
\usepackage{url}
\usepackage{amsmath}
\usepackage{amssymb}
\usepackage{bm}   
\usepackage{bbm}   
\usepackage{verbatim}
\usepackage{stmaryrd}
\usepackage{amsthm}

  \usepackage[dvips]{graphicx}

\theoremstyle{plain}    
\theoremstyle{plain}    
\theoremstyle{plain} 	\newtheorem{Cor}{Corollary}
\theoremstyle{plain} 	
\theoremstyle{plain} 	\newtheorem{The}{Theorem}
\theoremstyle{plain} 	
\theoremstyle{plain} 	
\theoremstyle{plain} 	
\theoremstyle{plain} 	
\theoremstyle{plain}	
\theoremstyle{plain}	 
\theoremstyle{plain}	

\input{cmechabbrev}

\begin{document}

\title{Time's Barbed Arrow:\\
Irreversibility, Crypticity, and Stored Information}

\author{James P. Crutchfield}
\email{chaos@cse.ucdavis.edu}
\affiliation{Complexity Sciences Center and Physics Department,
University of California at Davis, One Shields Avenue, Davis, CA 95616}
\affiliation{Santa Fe Institute, 1399 Hyde Park Road, Santa Fe, NM 87501}

\author{Christopher J. Ellison}
\email{cellison@cse.ucdavis.edu}
\affiliation{Complexity Sciences Center and Physics Department,
University of California at Davis, One Shields Avenue, Davis, CA 95616}

\author{John R. Mahoney}
\email{jrmahoney@ucdavis.edu}
\affiliation{Complexity Sciences Center and Physics Department,
University of California at Davis, One Shields Avenue, Davis, CA 95616}

\date{\today}

\bibliographystyle{unsrt}

\begin{abstract}
We show why the amount of information communicated between the past and
future---the \emph{excess entropy}---is not in general the amount of
information stored in the present---the \emph{statistical complexity}.
This is a puzzle, and a long-standing one, since the latter is what is
required for optimal prediction, but the former describes observed behavior.
We layout a classification scheme for dynamical systems and stochastic
processes that determines when these two quantities are the same or different.
We do this by developing closed-form expressions for the excess entropy in
terms of optimal causal predictors and retrodictors---the \eMs\ of
computational mechanics. A process's causal irreversibility and crypticity
are key determining properties.
\end{abstract}

\preprint{Santa Fe Institute Working Paper 09-XX-XXX}
\preprint{arxiv.org:09XX.XXXX [physics.gen-ph]}

\maketitle




Constructing a theory can be viewed as our attempt to extract from measurements
a system's hidden organization. This suggests a parallel with cryptography
whose goal \cite{Shan49a} is to not reveal internal correlations within an
encrypted data stream, even though it contains, in fact, a message. This is
essentially the circumstance that confronts a scientist when building a model
for the first time.

In this view, the now-long history in nonlinear dynamics to reconstruct models
from time series
\cite{Kant06a,Spro03a}
concerns
the case of \emph{self-decoding} in which the information used to build a model
is only that available in the observed process. That is, no ``side-band''
communication, prior knowledge, or disciplinary assumptions are allowed.
Nature speaks for herself only through the data she willingly gives up.

Here we show that the parallel is more than metaphor: building a model
corresponds directly to decrypting the hidden state information in measurements.
The results show why predicting and modeling are, at one and the same time,
distinct and intimately related. Along the way, a number of persistent
confusions about the role of (and different kinds of) information in prediction
and modeling are clarified. We show how to measure the degree of hidden
information and, along the way, identify a new kind of statistical
irreversibility that plays a key role.

Any process $\Prob(\Past,\Future)$ is a \emph{communication channel}:
It transmits information from the \emph{past}
$\Past = \ldots \MeasSymbol_{-3} \MeasSymbol_{-2} \MeasSymbol_{-1}$ to the
\emph{future} $\Future = \MeasSymbol_0 \MeasSymbol_1 \MeasSymbol_2 \ldots$
by storing it in the present. Here $\MeasSymbol_t$ is the random variable for
the measurement outcome at time $t$. Our goal is also simply stated: We wish to
predict the future using information from the past. At root, a prediction is
probabilistic, specified by a distribution of possible futures $\Future$
given a particular past $\past$: $\Prob(\Future|\past)$. At a minimum, a
good predictor needs to capture \emph{all} of the information $I$ shared
between past and future: $\EE = I[\Past;\Future]$---the process's
\emph{excess entropy} \cite[and references therein]{Crut01a}.

Consider now the goal of modeling---to build a representation that not only
allows good prediction, but also expresses the mechanisms that produce a
system's behavior. To build a model of a structured process (a channel),
computational mechanics \cite{CompMechMerge} introduced an equivalence
relation $\past \sim \past^\prime$ to group all histories that give rise to the
same prediction---resulting in a map from pasts to the \emph{causal states}:
$\epsilon(\past) =
  \{ \past^\prime: \Prob(\Future|\past) = \Prob(\Future|\past^\prime) \}$. 
A process's causal states, $\CausalStateSet = \Prob(\Past,\Future) / \sim$,
partition the space $\AllPasts$ of pasts into sets that are predictively
equivalent. The set of causal states can be discrete, fractal, or continuous.
State-to-state transitions are
denoted by matrices $T_{\CausalState \CausalState^\prime}^{(x)}$ whose elements
give the probability of transitioning from one state $\CausalState$ to the
next $\CausalState^\prime$ on seeing measurement value $\meassymbol$. The
resulting model, consisting of the
causal states and transitions, is called the process's \emph{\eM}.
 
Causal states have the Markovian property that they render the past and future
statistically independent; they \emph{shield} the future from the past
\cite{CompMechMerge}:
$\Prob(\Past,\Future|\CausalState)
  = \Prob(\Past|\CausalState) \Prob(\Future|\CausalState)$.
In this way, the causal states give a structural decomposition of
the process into conditionally independent modules.
Moreover, they are optimally predictive \cite{CompMechMerge} in the sense that
knowing which causal state a process is in is just as good as having the
entire past: $\Prob(\Future|\CausalState) = \Prob(\Future|\Past)$. In other
words, causal shielding is equivalent to the fact \cite{CompMechMerge} that the
causal states capture all of the information shared between past and future:
$I[\CausalState;\Future] = \EE$.

Out of all optimally predictive models $\PrescientStateSet$---for which
$I[\PrescientState;\Future] = \EE$---the \eM\ captures the minimal amount of
information that a process must store in order to communicate all of the excess
entropy from the past to the future. This is the \emph{statistical complexity}
\cite{CompMechMerge}: $\Cmu \equiv H[\CausalState] \leq H[\PrescientState]$.
In short, $\EE$ is the information transmission rate of the process, viewed
as a channel, and $\Cmu$ is the sophistication of that channel.

In addition to $\EE$ and $\Cmu$, another key (and historically prior)
invariant for dynamical systems and stochastic processes is the entropy rate
$\hmu$ which is the per-measurement rate at which the process generates
information---its degree of intrinsic randomness \cite{Shan48a}.
Importantly, the \eM\  immediately gives two of these three important
invariants: a process's
rate ($\hmu$) of producing information and the amount ($\Cmu$) of historical
information it stores in doing so.

To date, $\EE$ cannot be as directly calculated or estimated as the entropy
rate and the statistical complexity. This is truly unfortunate, since excess
entropy, and related mutual information quantities, are widely used diagnostics
for processes, having been applied to detect the presence of organization in
dynamical systems \cite{Fras86a,Casd91a,Spro03a,Kant06a}, in spin systems
\cite{Crut97a,Erb04a}, in neurobiological systems \cite{Tono94a,Bial00a}, and
even in language, to mention only a few applications. For example, in natural
language the excess entropy appears to diverge as
$\EE \propto L^{1/2}$, reflecting the long-range and strongly nonergodic
organization necessary for human communication \cite{Ebel94c,Debo08a}.

This state of affairs has been
a major impediment to understanding the relationships between modeling and
predicting and, more concretely, the relationships between (and even the
interpretation of) a process's basic invariants---$\hmu$, $\Cmu$, and $\EE$.
Here we clarify these issues by deriving explicit expressions for $\EE$ in
terms of the \eM, providing a unified information-theoretic analysis of
general processes.

The above development of \eMs\ concerns using the past to predict the future.
But what about retrodicting, using the future to retrodict the past?
Usually, one thinks of successive measurements occurring as time
increases. Now, consider scanning the measurement variables not in
the forward time direction, but in the reverse. The computational mechanics
formalism is essentially unchanged, though its meaning and notation need
to be augmented.

With this in mind, the previous mapping from pasts to causal states is
denoted $\FutureEps$ and it gave, what we will call, the
\emph{predictive} causal states
$\FutureCausalStateSet$. When scanning in the reverse direction, we
have a new relation, $\future \PastSim \future^\prime$, which groups futures
that are equivalent for the purpose of retrodicting the past:
$\PastEps(\future) =
  \{ \future^\prime: \Prob(\Past|\future) = \Prob(\Past|\future^\prime) \}$.
It gives the \emph{retrodictive} causal states
$\PastCausalStateSet = \Prob(\Past,\Future) / \PastSim$.
And, not surprisingly, we must also distinguish a process's forward-scan
\eM\ $\FutureEM$ from its reverse-scan \eM\ $\PastEM$. They assign
corresponding entropy rates, $\Futurehmu$ and $\Pasthmu$, and
statistical complexities, $\FutureCmu \equiv H[\FutureCausalState]$
and $\PastCmu \equiv H[\PastCausalState]$,
respectively, to the process.

Now we are in a position to ask some questions. Perhaps the most obvious is,
In which time direction is a process most predictable? The answer is that a
stationary process is equally predictable in either \cite{CompMechMerge}:
$\Pasthmu = \Futurehmu$. Somewhat surprisingly, though, the effort involved
in doing so is not the same \cite{Crut91bc}: $\PastCmu \neq \FutureCmu$.
Naturally, $\EE$ is mute on this score, since the mutual information $I$ is
symmetric in its variables \cite{Crut01a}.

The relationship between predicting and retrodicting a process, and ultimately
$\EE$'s role, requires teasing out how the states of the forward and reverse
\eMs\ capture information from the past and the future. To do this we must
analyze a four-variable mutual information:
$I[\Past;\Future;\PastCausalState;\FutureCausalState]$.
A large number of expansions of this quantity are possible. A systematic
development follows from Ref. \cite{Yeun91a} which showed that Shannon entropy
$H[\cdot]$ and mutual information $I[\cdot;\cdot]$ form a measure over
the space of events.

\begin{figure}[th]
\begin{center}
\resizebox{!}{1.7in}{\includegraphics{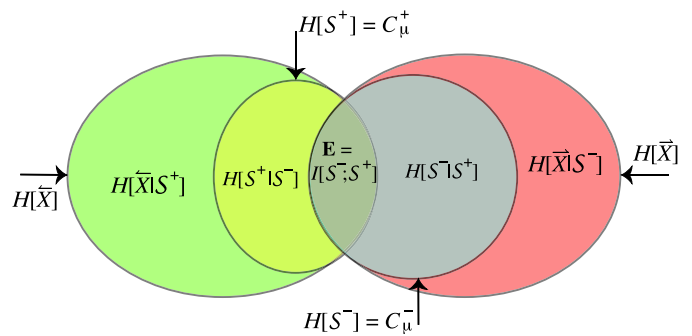}}
\end{center}
\caption{
  \EM\ information diagram for stationary hidden stochastic processes.
  }
\label{fig:eMIDiagram}
\end{figure}

Using an information diagram expansion, it turns out there are 15 possible
relationships to consider for
$I[\Past;\Future;\PastCausalState;\FutureCausalState]$. Fortunately,
this greatly simplifies in the case of using an \eM\ to represent
a process: There are only five relationships.
(See Fig. \ref{fig:eMIDiagram}.)
Simplified in this way, we are left with our main results which, due to the
preceding effort, are particularly transparent.
\begin{The}
Excess entropy is the mutual information between the predictive
and retrodictive causal states:
\begin{equation}
\EE = I[\PastCausalState;\FutureCausalState] ~.
\end{equation}
\label{EasCausalMI}
\end{The}
\vspace{-0.3in}
\noindent
Notably, the process's channel capacity $\EE = I[\Past;\Future]$ is the same as
that of the ``channel'' between the forward and reverse \eM\ states. Moreover,
the \emph{predictive statistical complexity} is given by
$\FutureCmu = \EE + H[\FutureCausalState|\PastCausalState]$
and the \emph{retrodictive statistical complexity} by
$\PastCmu = \EE + H[\PastCausalState|\FutureCausalState]$.

Theorem \ref{EasCausalMI} and its two companion results give an explicit
connection between
a process's excess entropy and its causal structure---its \eMs. More generally,
the relationships directly tie mutual information measures of observed
sequences to a process's structure. They will allow us to probe the properties
that control how closely observed statistics reflect a process's internal
hidden structure; that is, the degree to which observed behavior directly
reflects internal state information.

At this point we have two separate \eMs, one for predicting and one for
retrodicting. We will now show that one can do better, by combining causal
information from the past and future. Consider scanning a realization,
$\biinfinity = \past_t \future_t$, of the process in the forward
direction---seeing histories $\past_t$ and noting the series of causal states
$\FutureCausalState_t = \FutureEps(\past_t)$. Now change direction. What
reverse causal state is one in? This is
$\PastCausalState = \PastEps(\future_t)$. We describe the process of changing
scan direction with the \emph{bidirectional machine} $\BiEM$, which is given
by the equivalence relation $\BiEquiv$:
\begin{align*}
\BiEps(\biinfinity) & = \BiEps(\past,\future) \\
  & = \{ (\pastprime,\futureprime):
  	\pastprime \in \FutureEps(\past) ~\mathrm{and}~
	\futureprime \in \PastEps(\future) \}
\end{align*}
and has causal states $\BiCausalStateSet = \Prob(\Past,\Future) /
\BiEquiv \subset \FutureCausalStateSet \times \PastCausalStateSet$.
That is, the bidirectional causal state the process is in at time $t$ is
$\BiCausalState_t = (\FutureEps(\past_t),\PastEps(\future_t))$.
The amount of stored information needed to optimally predict \emph{and}
retrodict a process is $\BiEM$'s statistical complexity:
$\BiCmu \equiv H[\BiCausalState] = H[\PastCausalState,\FutureCausalState]$.

From the immediately preceding results we obtain the following simple, useful
relationship: $\EE = \FutureCmu + \PastCmu - \BiCmu$.
This suggests a wholly new interpretation of the excess entropy---in
addition to the original three reviewed in Ref. \cite{Crut01a}: $\EE$ is
exactly the difference between these statistical complexities. Moreover,
only when $\EE = 0$ does $\BiCmu = \FutureCmu + \PastCmu$.
The bidirectional machine is also efficient:
$\BiCmu \leq \FutureCmu + \PastCmu$.
And we have the bounds:
$\FutureCmu \leq \BiCmu$ and $\PastCmu \leq \BiCmu$.
These results say that taking into account causal information from the
past and the future is more efficient than ignoring one or the
other and than ignoring their relationship.

We noted above that predicting and retrodicting may require different amounts
of information storage ($\FutureCmu \neq \PastCmu$). It is helpful to use
\emph{causal irreversibility} to measure this asymmetry \cite{Crut91bc}:
$\CI \equiv \FutureCmu - \PastCmu$. With the above results, however, we see that
$\CI = H[\FutureCausalState|\PastCausalState] -
H[\PastCausalState|\FutureCausalState]$.
Note that irreversibility is also not controlled by $\EE$, as the
latter is scan-symmetric.

The relationship between excess entropy and statistical complexity established
by Thm. \ref{EasCausalMI} indicates that there are fundamental limitations on
the amount of a process's stored information ($\BiCmu$) directly present in
observations ($\EE$). We now introduce a measure of this: A process's
\emph{crypticity} is
$d(\FutureEM,\PastEM) \equiv H[\FutureCausalState|\PastCausalState]
+ H[\PastCausalState|\FutureCausalState]$.
This is the distance between a process's forward and reverse \eMs\ and
expresses most explicitly the difference between prediction and modeling.
To see this, we need the following connection.

\begin{Cor}
$\BiEM$'s statistical complexity is:
\begin{equation}
\BiCmu = \EE + d(\FutureEM,\PastEM) ~.
\end{equation}
\end{Cor}
\vspace{-0.05in}
\noindent
Referring to $d$ as crypticity derives from this result: It
is the amount of internal state information ($\BiCmu$) not directly
present in the observed sequence ($\EE$). That is, a process hides
$d$ bits of information.

If crypticity is low ($d \approx 0$), then much of the stored
information is present in observed behavior: $\EE \approx \BiCmu$. However,
when a process's crypticity is high, $d \approx \BiCmu$, then little of
it's structural information is directly present in observations. Moreover,
there are truly cryptic processes ($\EE \approx 0$) that are highly
structured ($\BiCmu \gg 0$). Little or nothing can be learned from measurements
about such processes's hidden organization.

\begin{figure}[th]
\begin{center}
\resizebox{!}{3.75in}{\includegraphics{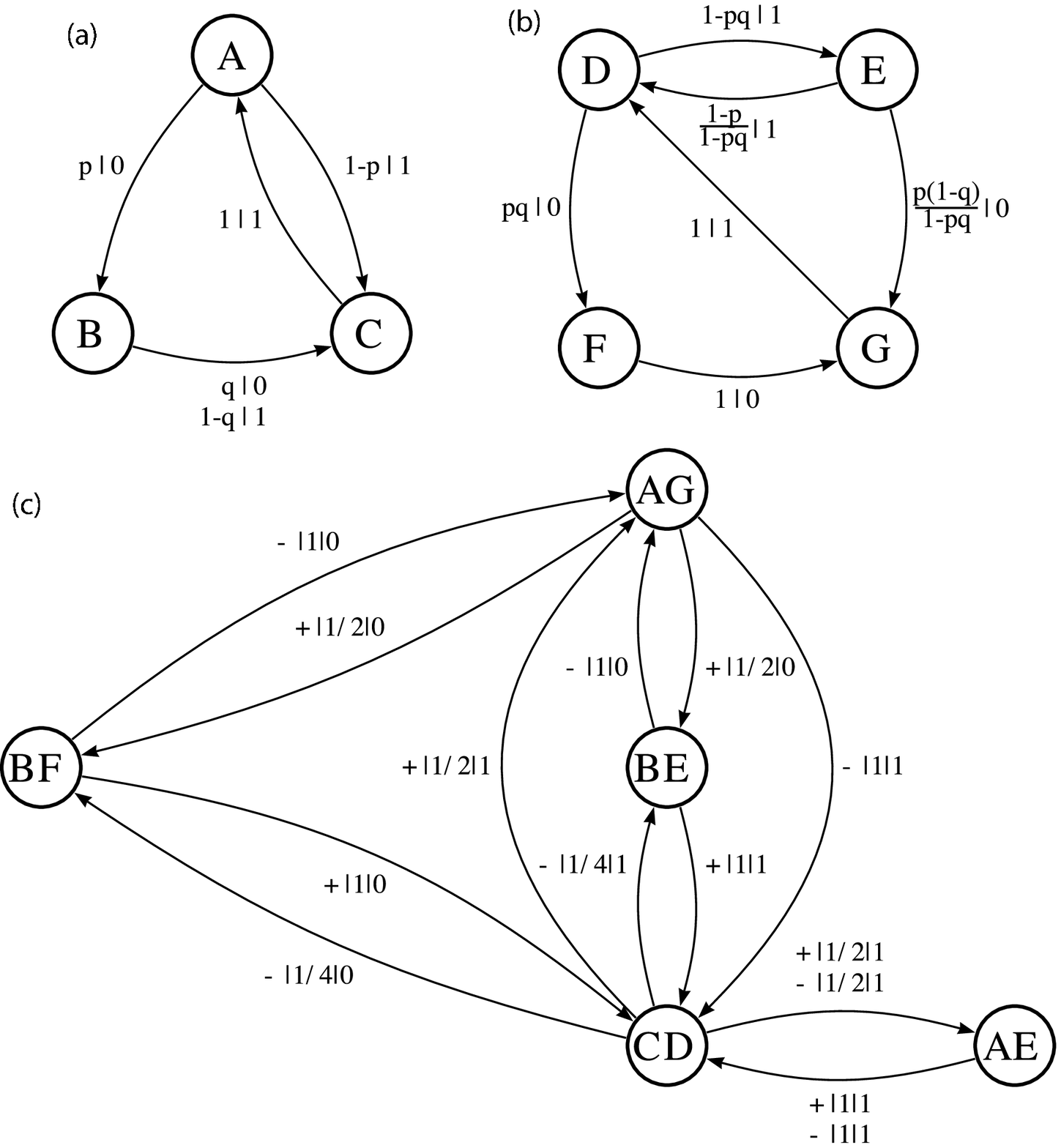}}
\end{center}
\caption{
  Forward and reverse \eMs\ for the RIP: $\FutureEM$ (a) and $\PastEM$ (b).
  Edge labels $t|x$ give the transition probabilities
  $t = T_{\CausalState \CausalState^\prime}^{(x)}$.
  The bidirectional machine $\BiEM$ (c) for $p = q = 1/2$.
  Edge labels here prepend the scan direction $\{-,+\}$.
  }
\label{fig:RIP}
\end{figure}

The \eM\ information diagram of Fig. \ref{fig:eMIDiagram} encapsulates all
of these results concisely. The diagram shows the key relationships between
information production ($H[\Future|\PastCausalState]$ and
$H[\Past|\FutureCausalState]$), excess entropy ($\EE = I[\Past;\Future]$), and
stored information ($\FutureCmu$ and $\PastCmu$). Analyzing the $4$-variable
information diagram showed that there are only four convex sets of interest.
These are depicted as differently shaded ellipses. $H[\Past]$ and $H[\Future]$
(two largest ellipses) are the entropies of the past and future, respectively,
which are the process's total information production. The information stored
in the predictive \eM\ $\FutureEM$ is its statistical complexity:
$\FutureCmu \equiv H[\FutureCausalState]$ (small ellipse on left); likewise
for $\PastEM$, $\PastCmu \equiv H(\PastCausalState)$ (small ellipse on right).
The excess entropy $\EE$ is the intersection of these sets; while the
statistical complexity $\BiCmu$ of the bidirectional machine $\BiEM$ is their
union; the crypticity $d(\FutureEM,\PastEM)$, their symmetric difference; and
their signed difference, the causal irreversibility $\CI$.

Consider an example that illustrates the typical process---cryptic
and causally irreversible. This is the random insertion process (RIP) which
generates a random bit with bias $p$. If that bit was a $1$, then it outputs
another $1$. If the random bit was a $0$, however, it inserts another random
bit with bias $q$, followed by a $1$.

Its forward \eM, see Fig. \ref{fig:RIP}(a), has three recurrent causal states
$\FutureCausalStateSet = \{ A, B, C \}$ and the transition matrices given there.
Figure \ref{fig:RIP}(b) gives $\PastEM$ which has four recurrent causal states
$\PastCausalStateSet = \{ D, E, F, G \}$. We see that the \eMs\ are not the
same and so the RIP is causally irreversible. A direct calculation gives
$\Prob(\FutureCausalState) = \Prob(A,B,C) = (1, p, 1) / (p+2)$ and
$\Prob(\PastCausalState) = \Prob(D,E,F,G) = (1, 1-pq, pq, p) / (p+2)$. If
$p = q = 1/2$, for example, these give us $\FutureCmu \approx 1.5219$ bits,
$\PastCmu \approx 1.8464$ bits, and $\hmu = 3/5$ bits per measurement.
The causal irreversibility is $\CI \approx 0.3245$ bits.

Let's analyze its bidirectional machine; shown in Fig.
\ref{fig:RIP}(c) for $p = q = 1/2$. The interdependence between the forward
and reverse states is given by:
\begin{align*}
& \Pr(\FutureCausalState, \PastCausalState) = \frac{1}{(p+2)(1-pq)} \\
& \times \bordermatrix{
   & D & E & F & G \cr
 A & 0 & 0 & pq(1-p) & p(1-pq) \cr
 B & 0 & (1-pq)^2 & p^2 q (1-q) & 0 \cr
 C & 1-pq & 0 & 0 & 0
} ~.
\end{align*}
By way of demonstrating the exact analysis now possible, $\EE$'s closed-form
expression for the RIP family is
\begin{align*}
\EE = \log_2 (p+2) - \frac{p \log_2 p }{p + 2} -
\frac{1-pq}{p + 2} H\left( \frac{1-p}{1-pq}\right) ~,
\end{align*}
where $H(\cdot)$ is the binary entropy function. The first two terms
on the RHS are $\FutureCmu$ and the last is
$H[\FutureCausalState|\PastCausalState]$.


Setting $p = q = 1/2$, one calculates that
$\Prob(\BiCausalState) = \Prob(AE,AG,BE,BF,CD) = \left(1/5,1/5,1/10,1/10,2/5\right)$.
This and the joint distribution give
$\BiCmu = H[\BiCausalState] \approx 2.1219$ bits,
but an $\EE = I[\FutureCausalState;\PastCausalState] = 1.2464$ bits. That is,
the excess entropy (the apparent information) is substantially less than the
statistical complexities (stored information)---a rather cryptic process:
$d \approx 0.8755$ bits.

To close, the main results establish that when $d > 0$ one cannot simply use
sequence information directly to represent a process as storing $\EE$ bits
of information. We must instead store $\Cmu$ bits of information, building a
causal model of the hidden state information.
Why? Because typical processes encrypt their state information within their
observed behavior. More precisely, observed information can be arbitrarily
small ($\EE \approx 0$) compared to the stored information ($\Cmu$).

In deriving an explicit relationship between excess entropy and the \eM, the
framework puts prediction on an equal footing with modeling and so allows for
a direct comparison between them. Also, as we demonstrated with the RIP example,
it gives a way to develop closed-form expressions for $\EE$.  Finally and most
generally, it reveals an intimate connection between unpredictability,
irreversibility, crypticity, and information storage.

Practically, the results clear up persistent confusions in several literatures
that conflate observed (mutual) information and a process's stored information.
Analyzing a process only in terms of mutual information misses
an arbitrarily large amount of a process's structure. When this happens, one
concludes that a process is more random than it is and that it has little
structure, when neither is true.

Chris Ellison was partially supported on a GAANN fellowship. The Network
Dynamics Program funded by Intel Corporation also partially supported this
work.

\vspace{-0.1in}
\bibliography{ref,chaos}

\end{document}

%% file: cmechabbrev.tex
\newcommand{\eM}     {$\epsilon$-machine}
\newcommand{\eMs}    {$\epsilon$-machines}
\newcommand{\EM}     {$\epsilon$-Machine}

\newcommand{\MeasSymbol}   { {X} }
\newcommand{\meassymbol}   { {x} }

\newcommand{\biinfinity}	{ \overleftrightarrow {\meassymbol} }
\newcommand{\Past}	{ \overleftarrow {\MeasSymbol} }
\newcommand{\past}	{ {\overleftarrow {\meassymbol}} }
\newcommand{\pastprime}	{ {\past}^{\prime}}
\newcommand{\Future}	{ \overrightarrow{\MeasSymbol} }
\newcommand{\future}	{ \overrightarrow{\meassymbol} }
\newcommand{\futureprime}	{ {\future}^{\prime}}

\newcommand{\AllPasts}	{ { \overleftarrow {\rm {\bf \MeasSymbol}} } }

\newcommand{\CausalState}	{ \mathcal{S} }

\newcommand{\CausalStateSet}	{ \boldsymbol{\CausalState} }
\newcommand{\AlternateState}	{ {\cal R} }

\newcommand{\PrescientState}	{ \widehat{\AlternateState} }

\newcommand{\PrescientStateSet}	{ \boldsymbol{\PrescientState}}

\newcommand{\Prob}		{ {\rm P}}

\newcommand{\Cmu}		{ {C_\mu}}
\newcommand{\hmu}		{ {h_\mu}}
\newcommand{\EE}		{ {\bf E}}

\newcommand{\CI}		{\Xi}

\newcommand{\FutureCausalState}	{ {\CausalState}^+ }

\newcommand{\PastCausalState}	{ {\CausalState}^- }

\newcommand{\BiCausalState}		{ {\CausalState}^{\pm} }

\newcommand{\FutureCausalStateSet}	{ {\CausalStateSet}^+ }
\newcommand{\PastCausalStateSet}	{ {\CausalStateSet}^- }
\newcommand{\BiCausalStateSet}	{ {\CausalStateSet}^{\pm} }
\newcommand{\eMachine}	{ M }
\newcommand{\FutureEM}	{ {\eMachine}^+ }
\newcommand{\PastEM}	{ {\eMachine}^- }
\newcommand{\BiEM}		{ {\eMachine}^{\pm} }
\newcommand{\BiEquiv}	{ {\sim}^{\pm} }
\newcommand{\Futurehmu}	{ h_\mu^+ }
\newcommand{\Pasthmu}	{ h_\mu^- }
\newcommand{\FutureCmu}	{ C_\mu^+ }
\newcommand{\PastCmu}	{ C_\mu^- }
\newcommand{\BiCmu}		{ C_\mu^{\pm} }
\newcommand{\FutureEps}	{ \epsilon^+ }
\newcommand{\PastEps}	{ \epsilon^- }
\newcommand{\BiEps}	{ \epsilon^{\pm} }

\newcommand{\PastSim}	{ \sim^- }